\documentclass[a4paper,twocolumn,11pt,unpublished]{quantumarticle}
\pdfoutput=1
\usepackage[utf8]{inputenc}
\usepackage[english]{babel}
\usepackage[T1]{fontenc}
\usepackage{amsmath}
\usepackage{hyperref}
\usepackage{mathtools}
\usepackage{comment}
\usepackage{tikz}
\usepackage{lipsum}
\usepackage{svg}

\usepackage{amssymb}
\usepackage{physics}
\usepackage{braket}
\usepackage{graphicx,subcaption}
\usepackage{bm}
\usepackage{subcaption}
\DeclareCaptionLabelFormat{custom}{Fig~\thefigure\alph{subfigure}}
\begin{document}

\title{In Search of Lost Tunneling Time}

\author{Pablo M. Maier}
\email{maier@mbi-berlin.de}
\affiliation{Max Born Institute for Nonlinear Optics and Short Pulse Spectroscopy, Max-Born-Straße 2A, 12489 Berlin, Germany}


\author{Serguei Patchkovskii}
\affiliation{Max Born Institute for Nonlinear Optics and Short Pulse Spectroscopy, Max-Born-Straße 2A, 12489 Berlin, Germany}

\author{Misha Yu. Ivanov}
\affiliation{Max Born Institute for Nonlinear Optics and Short Pulse Spectroscopy, Max-Born-Straße 2A, 12489 Berlin, Germany}
\affiliation{Humboldt-Universität zu Berlin, Unter den Linden 6, 10117 Berlin, Germany}
\affiliation{Technion, Haifa, 32000, Israel}

\author{Olga Smirnova}
\affiliation{Max Born Institute for Nonlinear Optics and Short Pulse Spectroscopy, Max-Born-Straße 2A, 12489 Berlin, Germany}
\affiliation{Technische Universität Berlin, Straße des 17. Juni 135, 10623 Berlin, Germany}
\affiliation{Technion, Haifa, 32000, Israel}

\maketitle

\begin{abstract}
    The measurement of tunneling times in strong-field ionization has been the topic of much controversy in recent years, with the attoclock and Larmor clock being two of the main contenders for correctly reproducing these times. While the non-zero Larmor tunneling time has been unambiguously detected in Ref.\cite{ramos2020measurement}, the tunneling time measured by the attoclock appears to vanish in tunnel ionization of a hydrogen atom \cite{sainadh2019attosecond}. By expressing the attoclock as the weak value of temporal delay \cite{sokolovski2018no}, we extend its meaning beyond the traditional setup 
    and connect it to the Steinberg weak-value interpretation of the Larmor clock \cite{steinberg1995much}. 
    Our approach allows us to derive the position-resolved attoclock tunneling time.  We show how this time, while non-zero at the tunnel exit, vanishes at the detector, far away from the atom. Formally, this means that the attoclock does not measure the “local” Larmor time detected in Ref.\cite{ramos2020measurement}, but instead a “non-local” time closely related to the phase time.
\end{abstract}

Tunneling time  refers to the time it takes for a quantum particle to tunnel through a potential barrier \cite{landauer1994barrier}. Since the pioneering work of MacColl \cite{MacColl_pra1932} and Hartman \cite{hartman1962tunneling}, the extensive literature on this topic has continually challenged our understanding of time. The exploration of tunneling time offers a journey that rivals, in depth and intensity, the introspective odyssey described by Proust in his celebrated work devoted to the essence of time.

This journey begins with the absence of a self-adjoint time operator conjugate to the Hamiltonian. Such an operator, if it existed, could provide a foundation for defining fundamentally quantum observables like tunneling time \cite{pauli2012general}.

At this point, since there is no unique way to define this time, the journey splits into several ways, just like one learns about "Swan's way" and "Guermant's way" in Proust's work. Each way emphasizes different aspects of tunneling by introducing an operational definition, associated with detection of different physical observables. This has led to qualitatively different results. For example, some approaches yield tunneling times that are zero (or close to zero) \cite{torlina2015interpreting,pollak2017book, ni_rost2016, eicke_lein2018, sainadh2019attosecond,flores2024}, while others yield  non-zero tunneling times \cite{landsman2014ultrafast,steinberg1995much, camus2017,klaiber2022}. 
The  most apparent recent controversy is related to two different operational definitions or measurement protocols, known as the Larmor clock and the attoclock.

The attoclock \cite{eckle2008attosecond} explores tunneling of an electron from an atom through the barrier created by an intense low-frequency field, a process known as optical tunneling \cite{keldysh2024ionization}. Utilizing ultra-short pulses of nearly circularly polarized light, the attoclock maps the tunneling time onto the angular distribution of photoelectrons. The primary observable is the angle between the orientation of the maximal value of light polarization vector and the orientation of the most probable final momentum of the electron ( see reviews \cite{landsman_review2015, Satya_Sainadh2020}.) 

In the case of tunneling from a short-range potential and in the absence of tunneling delay, this angle is 90 degrees. Any deviation from this value is referred to as the off-set angle. The off-set angle encodes the scattering phase \cite{torlina2015interpreting}, and potentially carries information about the tunneling time \cite{eckle2008attosecond}. In the case of a very thin barrier, the offset angle can become non-negligible—even for a short-range potential—due to the interference between the tunneled trajectory and the trajectory that was first reflected at the exit point of the barrier, then reflected a second time from the core, and finally tunneled again \cite{klaiber2022}.

The time measured by the attoclock is derived from the offset angle using the relation $\tau_{A} = \theta_{\rm off}/\omega$, where $\omega$ represents the frequency of the ionizing field. The tunneling time is obtained by subtracting the time associated with the scattering phase from the total time measured by the attoclock: $\tau_{\rm A}^{\rm tun}=\tau_{\rm A}-\tau_{\rm A}^{\rm scat}$ \cite{torlina2015interpreting}. 

In the Larmor clock \cite{rybachenko1967time, baz1966lifetime, baz1967quantum}, a magnetic field localized within the barrier region interacts with the spin of the tunneling particle, recording the tunneling time via the spin precession angle. The tunneling time measured by the Larmor clock is determined from the precession angle using the relation $\tau_{\rm L}^{\rm tun} = \phi_{\rm L}/\Omega_{\rm L}$, where $\Omega_{\rm L}$ denotes the Larmor frequency.

Just as "Swann's Way" and "Guermantes Way" offer different perspectives and observables in Proust's narrative, the observables associated with the "Larmor's way" and "Attoclock's way" also differ significantly.

The disparity between these approaches reflects two distinct interpretations of time in quantum mechanics: as a parameter linked to the phase of a wavefunction, or as a measure of the number of particles with given velocity in a confined region of space relative to the total available space during a given observation period.

In the first interpretation, relevant to the \emph{Attoclock's way}, time does not need to be local; it does not have to be defined at every point in space. In contrast, the second interpretation, pertinent to the \emph{Larmor's way}, defines time locally and allows one to track how it evolves spatially \cite{steinberg1995much}.

However, there are also important similarities between these two perceptions of time. 
Spending time in a specific region may imply both entering and exiting it, resulting in a quantum state that differs between the beginning and the end of the process.
In both interpretations, determining such times requires access to the complex-valued \emph{transition amplitudes} of the underlying process, rather than probabilities alone. Consequently, the times become complex.

At this point both  the \emph{Attoclock's way} and \emph{Larmor's way}  intersect with two other important concepts. 

The first is the concept of time-resolved measurements, which has reached unprecedented precision through advancements in attosecond physics. This progress has enabled the full characterization of transition amplitudes using various setups, such as the attosecond streak camera, RABBIT, attoclock, and high harmonic spectroscopy \cite{krausz2009attosecond}. The latter techniques have facilitated the detection of complex-valued ionization times \cite{shafir2012resolving, pedatzur2015attosecond} related to optical tunneling.

The second concept is that of weak measurements \cite{aharonov1988result}, which provide an elegant framework for accessing complex-valued observables in quantum mechanical measurements. Weak measurements achieve this by mapping the real and imaginary components of a complex-valued observable onto distinct, real, and positive detector "clicks". This method was used by Steinberg to map the complex Larmor time, a result later realized experimentally \cite{ramos2020measurement}. A weak measurement is a "perturbative" measurement, and thus enables the interference between unperturbed and perturbed amplitudes, thereby allowing independent access to the real and imaginary parts of the interference term \cite{dressel2014colloquium}.


The connection between the  \emph{Attoclock's way} and the \emph{Larmor's way} was intuited by A. Landsmann \cite{landsman2014ultrafast} in an attempt to describe the outcome of the attoclock experiments \cite{landsman2014ultrafast} in terms of the Larmor times.
The Larmor time calculated for a triangular barrier yielded an excellent agreement with the attoclock experiment on Helim atoms \cite{landsman2014ultrafast} meaning that $\tau_{\rm A}^{\rm tun}=\tau_{\rm L}^{\rm tun}$. However, the ab-initio simulations on Helium by V. P. Majety and  A. Scrinzi \cite{majety2017absence}, stimulated by the experiment \cite{landsman2014ultrafast} had an opposite outcome: $\tau_{\rm A}^{\rm tun}=0$.

After this junction \cite{landsman2014ultrafast} the  \emph{Attoclock's way} and the \emph{Larmor's way} diverged to reveal new experiments.

Along the \emph{Attoclock's way}, an experiment by Camus et al in noble gases revealed the tunneling time \cite{camus2017}. However, a more recent experiment by Sainadh et al. \cite{sainadh2019attosecond} on a simpler system -- the hydrogen atom -- building on earlier theoretical predictions by Torlina et al. \cite{torlina2015interpreting}, failed to detect a tunneling time consistent with these predictions. Specifically, the results indicated that $\tau_{\rm A}^{\rm tun} = 0$, suggesting no measurable tunneling delay.



Along the \emph{Larmor's way}, a recent experiment by Ramos et al. realized the Larmor clock for electronic transitions in Bose-condensed Rb atoms tunneling through an optical barrier. This experiment detected a non-zero Larmor tunneling time, $\tau_{\rm L}^{\rm tun} \neq 0$.

Together, this body of work suggests that we still have an unsolved question about the relationship between the Larmor and attoclock times: $\tau_{\rm A}^{\rm tun}\overset{?}{=} \tau_{\rm L}^{\rm tun}$. Here we zoom into the similarities and  differences between the two times, apply them to the attoclock problem in its standard formulation making the first steps to address the question:
"Shall one expect to find the Larmor time in the attoclock measurement?". Pertinent work analyzing the nature of the tunneling delays in strong field ionization and their values at the exit of the tunneling barrier using the concepts of "virtual detector", Wigner trajectories,  Bohmian trajectories and Wigner or Husimi distributions includes \cite{czirjak2000wigner,ivanov2005anatomy,teeny_prl2016,teeny_pra2016,yakaboylu_pra2013,Douget_pra2018}.

\section{Theoretical Background}\label{sec: Larmor clock}
In this section, we review the foundational concepts underlying the Larmor clock and the Attoclock, emphasizing their theoretical frameworks and operational principles.

\subsection{Larmor Clock}
The core concept of the Larmor clock \cite{rybachenko1967time, baz1966lifetime, baz1967quantum} is to encode the tunneling time into the evolution of an additional degree of freedom that remains unaffected by the tunneling process, such as the electron's spin.
To this end, a magnetic field is confined to the barrier region. The spin of the tunneling particle precesses under the influence of the magnetic field. Since this precession only takes place inside the barrier, it can be used to extract the time $\tau^y_{\rm L}$ that passes during tunneling.  
Let's consider a potential barrier $V(x,t)$, with the classical turning points $a,b$ representing the entry and exit of the tunnel. The projector $\hat{\theta}_{\rm B}$ onto the barrier region $[a,b]$, can be written in terms of the Heaviside function $\theta(x)$:
\begin{align}
    \bra{x}\hat{\theta}_\text{B}\ket{x} = \theta(x - a) - \theta(x - b),
\end{align}
where $x$ is the degree of freedom involved in tunneling. The electron is described by the Pauli equation for a weak magnetic field:
\begin{align}
    i \frac{\partial \Psi}{\partial t} &= \Bigg[ \frac{\hat{p}^2}{2} + V(x,t)\\
    &\quad+ \frac{1}{2} Bf(t)\hat{\theta}_{\rm B}\cdot( \hat{L}_z + g \hat{S}_z ) \Bigg]\Psi\,, \label{eq: pauli}
\end{align}
where $B$ is the magnetic field strength, $g$ is the $g$-factor and $f(t)$ is the temporal envelope of the magnetic field. We use the ansatz $|\Psi\rangle=\ket{\phi(t)}|\chi(t)\rangle$, where $|\chi(t)\rangle$ represents the spin. The ansatz neglects the influence of the magnetic field on the tunneling dynamics, $\phi(x,t)$. To ensure that the magnetic field does not perturb the electron's dynamics, it must remain sufficiently weak. Consequently, its coupling to the electron's orbital angular momentum can also be neglected \cite{falck1988larmor}, leading to the following one-dimensional equation:
\begin{align}
    i \frac{\partial \Psi}{\partial t} = \left[ \frac{\hat{p}^2}{2} + V(x,t) + \Omega_{\rm L} f(t) \hat{\theta}_{\rm B} S_z \right]\Psi\,, \label{eq: pauli no orbit1}
\end{align}
where $\Omega_{\rm L}= \frac{1}{2} g B$ is the Larmor frequency. 

Since the magnetic field does not affect the tunneling degree of freedom, tunneling is fully described by the Hamiltonian $H=\frac{\hat{p}^2}{2} + V(x,t)$ with the corresponding solution of the TDSE being  $|\phi(t)\rangle$.

The spin degree of freedom, in turn, is not affected by the tunneling barrier and evolves in the weak magnetic field according to the propagator:
\begin{align}\label{eq:propagator}
e^{-i\int_{t_0}^{t}dtH_B(t')}\simeq 1-i\int_{t_0}^{t}dt' H_B(t'),
\end{align}
where $H_B\equiv\Omega_{\rm L} f(t)\hat{\theta}_{\rm B} S_z$:
\begin{align}
    |\chi(t)\rangle = |\chi(t_0)\rangle- i \Omega_{\rm L} F(t)\hat{\theta}_{\rm B} S_z|\chi(t_0)\rangle, \label{eq: pauli no orbit2}
\end{align}
where $F(t)=\int_{t_0}^{t}dt' f(t')$.
After tunneling, the electron is characterized by the wave-function $|\Psi\rangle=|\phi(t)\rangle|\chi(t)\rangle$. We can now isolate the transmitted electrons by projecting \(\Psi\) onto the final state corresponding to the transmitted electrons: $|\Psi_{\rm T}(t)\rangle=\hat{P}_{\rm T}|\Psi(t)\rangle$ using the projector  $\hat{P}_{\rm T}=|\phi_{\rm T}(t)\rangle\langle \phi_{\rm T}(t)|$, where $|\phi_{\rm T}(t)\rangle$ describes the corresponding boundary condition. Explicitly, 
\begin{align}\label{eq:larmor time angle1}
|\Psi_{\rm T}(t)\rangle=\Biggl[\langle \phi_{\rm T}(t) |\phi(t)\rangle-\\i\Omega_{\rm L} F(t)\langle \phi_{\rm T}(t)|\hat{\theta}_{\rm B}| \phi (t)\rangle S_z\Biggr]|\phi_{\rm T}(t)\rangle|\chi(t_0)\rangle.
\end{align}
The angle of spin precession for a spin oriented along the x-direction prior to tunneling is 
\begin{align}\label{eq:larmor time angle2}
\phi_{\rm L}^y = \arctan \frac{\braket{S_y}}{\braket{S_x}}\simeq \frac{\braket{S_y}}{\braket{S_x}}\,.
\end{align}
To measure this angle, it is necessary to measure the $\braket{S_y}$ and $\braket{S_x}$ spin components after tunneling at some time $t$. Since the magnetic field is weak, we only include the leading terms with respect to $B$. The expectation values of the operators $\hat{S}_x$ and $\hat{S}_y$ are:
\begin{align}
    \braket{\hat{S}_x} &= \frac{1}{2}\int dx \Psi_{\rm T}^\dagger \sigma_x \Psi_{\rm T} = \biggl|\langle \phi_{\rm T}(t) |\phi(t)\rangle \biggr|^2\,,
\end{align}
and
\begin{align}
    \braket{\hat{S}_y} = \frac{1}{2} \int dx \Psi_{\rm T}^\dagger \sigma_y \Psi_{\rm T} =\\ \Omega_{\rm L} F(t)\langle \phi_{\rm T}(t)|\hat{\theta}_{\rm B}| \phi (t)\rangle \langle \phi_{\rm T}(t) |\phi(t)\rangle\,.
\end{align}

The Larmor time is  
\begin{align} \label{eq:larmor time t}
    \tau^y_{\rm L} = \frac{\phi_{\rm L}^y}{\Omega_{\rm L}}=\text{Re}\frac{F(t)\langle \phi_{\rm T}(t)|\hat{\theta}_{\rm B}| \phi (t)\rangle }{\langle \phi_{\rm T}(t) |\phi(t)\rangle}\,.
\end{align}
The third component of spin $\braket{\hat{S}_z}$ gives access to the imaginary component of the Larmor time via $\phi_{\rm L}^z \simeq \frac{\braket{S_z}}{\braket{S_x}}$ :
\begin{align} \label{eq:larmor time Im}
    \tau^z_{\rm L} = \frac{\phi_{\rm L}^z}{\Omega_{\rm L}}=\text{Im}\frac{F(t)\langle \phi_{\rm T}(t)|\hat{\theta}_{\rm B}| \phi (t)\rangle }{\langle \phi_{\rm T}(t) |\phi(t)\rangle}\,.
\end{align}
In case of a stationary influx of the particles (plane waves) the total time of the experiment $F(t)$ can be written as the total amount of particles divided by the current $F(t)=\frac{N}{\dot{N}}$. Normalizing the total amount of the particles to 1, we obtain $F(t)=1/k$, where $k$ is the value of the current. Thus, in stationary regime for a continuous beam of particles the Larmor time can be written as:
\begin{align} \label{eq:larmor time stationary}
    \tau^y_{\rm L} =  \frac{\phi_{\rm L}^y}{\Omega_{\rm L}}=\text{Re}\frac{1}{k}\frac{\langle \phi_{\rm T}|\hat{\theta}_{\rm B}| \phi \rangle }{\langle \phi_{\rm T} |\phi\rangle}\,.
\end{align}
The additional simplifications come from the fact that all time-dependent phases are the same in numerator and denominator.

Two comments are in order.
First, Eqs.\,(\ref{eq:larmor time t},\ref{eq:larmor time stationary}) show that the  Larmor time is proportional to the matrix element of an operator $\hat{\theta}_{\rm B}$ corresponding to the transition from the evolved initial state $| \phi (t)\rangle$ to a different final state $|\phi_{\rm T}(t)\rangle$. Despite the fact that this matrix element is complex-valued, it can be accessed in the standard measurement, because the real and imaginary parts of the time are mapped into different real and positive observables \cite{steinberg1995much,ramos2020measurement}. 

Second, $\hat{\theta}_{\rm B} / k$ plays a role of the dwell time operator \cite{steinberg1995much}. This concept originates from the work of Sokolovski and Baskin \cite{sokolovski1987traversal}, which introduces the quantum traversal time, by generalizing a classical expression $t^{cl}_{\Gamma}=\int_{t_{1}}^{t_{2}}\Theta_{\Gamma}(\mathbf{r}(t))dt$, where $\Theta_{\Gamma}(\mathbf{r})=1$ inside the region $\Gamma$, and zero otherwise.
The generalization to the quantum case $t_{\Gamma}=\int_{t_{1}}^{t_{2}}\Theta_{\Gamma}(\mathbf{r}(\cdot))dt$ amounts to extending it to quantum trajectories, including all the multitude of Feynman paths $\mathbf{r}(\cdot)$. The Larmor time is calculated using Feynman path integrals and by expressing the time in the propagator \eqref{eq:propagator} via $t_{\Gamma}$, i.e. expressing the time via quantum trajectories. Following \cite{steinberg1995much}, we will use the dwell time operator $\hat{\theta}_{\rm B} / k$ to interpret the Larmor time as the outcome of a weak measurement.

\subsection{Weak Values and the Larmor Time}\label{sec: weak and larmor}
A weak measurement \cite{aharonov1988result} is a measuring protocol that gives access to the weak value of some operator $\hat{A}$:
\begin{align}\label{eq: weak value}
    A_\text{w} = \frac{\bra{\psi_\text{f}}\hat{A}\ket{\psi_\text{i}}}{\braket{\psi_\text{f} | \psi_\text{i}}}\,.
\end{align}

This is achieved through a von Neumann measurement, where the pointer of a measurement device becomes entangled with the system under observation via an interaction Hamiltonian \(\hat{H}_{\rm int}\). A post-selection on final states \(\ket{\psi_\text{f}}\) is then performed, allowing only a specific subset of measurement device states to be read out. 

The Larmor clock can be reformulated within the framework of weak measurement \cite{steinberg1995much} by using the operator of dwell time \(\hat{\theta}_{\rm B}/k\). Below, we follow the approach outlined in \cite{aharonov1988result} to express the Larmor time in the language of weak measurements \cite{steinberg1995much}.

The first step in a weak measurement is a von Neumann measurement. In this process, a system and a measurement device are prepared in an initial state. Taking the Larmor clock as an example, the system to be measured is the wavefunction of the tunneling electron, while the measurement device corresponds to the spin. The combined initial state of the spin and wavefunction can be expressed as:
\begin{align}
    \ket{\Psi_{\rm 1}} = \ket{+}_x \ket{\psi_{\rm i}}\,,
\end{align}
where \(\ket{+}_x\) represents the initial spin state along the \(x\)-axis, and \(\ket{\psi_{\rm i}}\) denotes the initial wavefunction of the tunneling electron.

Following this preparation, an interaction Hamiltonian \( H_{\rm int} \) acts on the system and the measurement device for a duration \(\tau\), entangling the two. For the Larmor clock, this interaction is given by \(\Omega_{\rm L} f(t) \hat{S}_z \otimes \hat{\theta}_{\rm B}\), where the notation is consistent with that introduced in the previous section. After this interaction, the state of the system evolves to:
\begin{align}
    \ket{\Psi_{\rm 2}} = e^{-i\Omega_{\rm L} F(t) \hat{S}_z \otimes \hat{\theta}_{\rm B}} \ket{+}_x \ket{\psi_{\rm i}}\,.
\end{align}
As discussed in the previous section, \(F(t)\) can be approximated by \(1/k\) for plane waves in a stationary process, bringing the dwell time operator \(\hat{\theta}_{\rm B}/k\) into play.

In a standard von Neumann measurement, this would mark the final step, followed by a projective measurement on the spin, i.e., the measurement device. In contrast, a weak measurement modifies this procedure by introducing a projection onto a specific final system state \(\ket{\psi_{\rm f}}\), which represents a particular sub-ensemble of measurement outcomes. This projection acts as a post-selection step. In the Larmor clock setup, this post-selection is realized by measuring only particles that have successfully traversed the potential barrier. Following this protocol, the state of the system becomes:
\begin{align}
    \ket{\Psi_3} &= \ket{\psi_{\rm f}}\bra{\psi_{\rm f}}e^{-i\frac{\Omega_{\rm L}}{k}\hat{S}_z\otimes \hat{\theta}_{\rm B}}\ket{+}_x\ket{\psi_{\rm i}}\,\\
    &= \frac{\ket{\psi_{\rm f}}\bra{\psi_{\rm f}}}{\sqrt{2}}\left(e^{-i\frac{\Omega_{\rm L}}{2k} \hat{\theta}_{\rm B}}\ket{+} + e^{i\frac{\Omega_{\rm L}}{2k} \hat{\theta}_{\rm B}}\ket{-}\right)\ket{\psi_{\rm i}}\,.
\end{align}
If the conditions of a weak measurement are met \cite{aharonov1988result}, we get 
\begin{align}
    \ket{\Psi_3} &\approx \frac{\ket{\psi_{\rm f}}\bra{\psi_{\rm f}}}{\sqrt{2}}\left(e^{-i\frac{\Omega_{\rm L}}{2k} {\theta}_{\rm B}^{\rm w}}\ket{+} + e^{i\frac{\Omega_{\rm L}}{2k} {\theta}_{\rm B}^{\rm w}}\ket{-}\right)\ket{\psi_{\rm i}}\,,
\end{align}
where ${\theta}_{\rm B}^{\rm w}$ is the weak value [see Eq.\,\eqref{eq: weak value}] of the projector onto the barrier.
On average a measurement of $\hat{S}_x$ and $\hat{S}_y$ gives:
\begin{align}
    \braket{\Psi_3|\hat{S}_x|\Psi_3} &= \frac{1}{2} |\braket{\psi_{\rm i}|\psi_{\rm f}}|^2 \cos(\Omega_{\rm L}{\theta}_{\rm B}^{\rm w} / k)\,,\\
    \braket{\Psi_3|\hat{S}_y|\Psi_3} &= \frac{1}{2} |\braket{\psi_{\rm i}|\psi_{\rm f}}|^2 \sin(\Omega_{\rm L}{\theta}_{\rm B}^{\rm w} / k)\,.
\end{align}
With Eq.\,\eqref{eq:larmor time angle2} and using $\Omega_{\rm L}{\theta}_{\rm B}^{\rm w}\ll1$ the Larmor time is
\begin{align}\label{eq: larmor time weak}
    \tau_{\rm L} \approx \frac{{\theta}_{\rm B}^{\rm w}}{k} = \frac{1}{k} \frac{\braket{\psi_{\rm f} | \hat{\theta}_{\rm B} | \psi_{\rm i}}}{\braket{\psi_{\rm f} | \psi_{\rm i}}}\,.
\end{align}
This implies that the Larmor time corresponds to the weak value of the dwell time operator. Notably, it is a self-contained concept, independent of the spin degree of freedom.

A brilliant realization of the Larmor clock measurement is demonstrated in the BEC experiment \cite{ramos2020measurement}. This experiment illustrates how weak values can be encoded in strong measurements of a conjugate degree of freedom and subsequently recovered. 

In this setup, two-level atoms tunnel through an optical barrier while a laser field, which creates the barrier, simultaneously drives transitions in the two-level system. The Hamiltonian governing the system is given by \( H = \Omega_R(x)\sigma_z \), where \(\Omega_R(x)\) is the (two-photon) Rabi frequency, effectively playing the role of the magnetic field which is strictly confined to the barrier region. The term \(\sigma_z\) represents the \(z\)-component of the Pauli matrix, corresponding to the pseudospin.

This system involves two degrees of freedom: the atom's position in real space, described by its center-of-mass wavefunction \(|\psi(x)\rangle\), and the electronic (pseudospin) wavefunction $|\chi(t,x)\rangle=(a(t), b(t)) $, expressed in the basis of the atom's two dressed states.

\subsection{Attoclock}\label{sec: attoclock}
In the attoclock setup \cite{eckle2008attosecond} an atom is ionized by a pulse of circularly polarized light. The electric field and the binding force of the atom create a rotating potential barrier through which the electron can tunnel. The angularly resolved photoelectron spectrum is then measured.

After the electron emerges from the tunneling barrier, its subsequent motion can be treated classically. For a short-range binding potential the conservation of canonical momentum \(\bm{p}\) provides a mapping between the ionization time \(t_{\rm i}\) and the photoelectron spectrum. This relationship is expressed as:
\begin{align}\label{eq: canonical}
    \bm{v}(t_{\rm i}) - \bm{A}(t_{\rm i}) = \bm{p}\,,
\end{align}
where \(\bm{v}(t_{\rm i})\) represents the velocity of the electron at the time of ionization, and \(\bm{A}(t_{\rm i})\) is the vector potential of the circularly polarized light at that instant.

This mapping can be used to associate a time \(t_{\rm i}\) with the peak of the photoelectron spectrum. For short-range binding potentials, the photoelectron spectrum can be computed using the Perelomov, Popov, and Terent’ev (PPT) approach \cite{perelomov1966ionization, perelomov1967ionization}. For the final momentum of the electron at the detector \(\bm{p} = (p, \theta)\)  the complex ionization time \(t_{\rm s} = t_{\rm i} + i \tau\) is \cite{barth2011nonadiabatic}:
\begin{align}
    &\omega t_{\rm i} = 2\pi N + \theta, \quad (N \in \mathbb{Z}) \label{eq:saddle solution} \\
    &\omega \tau = \text{arcosh}\left(\frac{A_0}{2p} \left[\left(\frac{p}{A_0}\right)^2 + \gamma^2 + 1\right]\right),
    \intertext{where}
    &\gamma = \sqrt{2I_{\rm p}}/A_0\,.
\end{align}

Equation \eqref{eq:saddle solution} illustrates how the angular offset \(\theta\) of the photoelectron distribution can be used to infer the time of ionization \(t_{\rm i}\). Symmetry arguments show that in the absence of the electron-core interaction, for short-range binding potentials, there is no angular offset (see \ref{sec:spectrum PPT}). However, for real atoms, such as hydrogen, Coulomb corrections \(\Delta t_{\rm i}\) to the short-range saddle-point time \(t_{\rm i}\) must be taken into account \cite{kaushal2013nonadiabatic, kaushal2015opportunities, torlina2013time, torlina2015interpreting}.


\subsection{Weak Values and the Attoclock}
In scattering, the weak value of temporal delay can be expressed as \cite{sokolovski2018no}:
\begin{align}
    \braket{t}_{\rm w}^{\rm scat} = \frac{\int_{t_0}^{t_{\rm f}} dt\, t\,\xi^{\rm scat}(t,E)}{T(E)}\,,
    \intertext{where the scattering amplitude}
    T(E) = \int_{t_0}^{t_{\rm f}} dt\, \xi^{\rm scat}(t,E)
\end{align}
is the integral over its temporal density $\xi^{\rm scat}(t,E)$. The real part of $\braket{t}_{\rm w}^{\rm scat}$ is the Eisenbud–Wigner–Smith time delay. Analogously, the ionization amplitude can be seen as the temporal integral
\begin{align}
    a_{\bm p} &= \braket{\bm{p}(t_{\rm f})|\psi(t_{\rm f})} = \int_{t_0}^{t_{\rm f}} dt\,\xi^{\rm ion}(t,\bm{p})
\end{align}
over the temporal density $\xi^{\rm ion}(t,\bm{p})$. The analytic R-matrix theory (see Ref.\,\cite{torlina2012time}) has shown that $\xi^{\rm ion}(t,\bm{p}) \propto e^{iI_{\rm p}(t-t_0)}$ and thus the weak value of temporal delay in ionization is (see \ref{sec:attoweak} for a detailed derivation)
\begin{align}
    \braket{t}_{\rm w}^{\rm ion} &= \frac{\int_{t_0}^{t_{\rm f}} dt\, t\,\xi^{\rm ion}(t,\bm{p})}{a_{\bm p}}\nonumber\\
    &= t_0 - i \frac{1}{a_{\bm p}}\left(\frac{\partial a_{\bm p}}{\partial I_{\rm p}}\right)_{\tilde{\kappa} = \text{const.}}\,.
    \label{eq:weak ion}
\end{align}
Applying the saddle-point approximation to the integral in Equation \eqref{eq:weak ion} gives
\begin{align}
    \braket{t}_{\rm w}^{\rm ion} &=t_0 - i \frac{1}{a_{\bm p}}\left(\frac{\partial a_{\bm p}}{\partial I_{\rm p}}\right)_{\tilde{\kappa} = \text{const.}} = t_{\rm s}\,.\label{eq: saddle time}
\end{align}

It was demonstrated in Ref.\,\cite{torlina2015interpreting} that the attoclock measures the real part, \(\Re t_{\rm s}\), of the saddle-point time. This result was re-derived using an alternative approach in Ref.\,\cite{kaushal2015opportunities} and further verified through comparison with ab-initio calculations for hydrogen in Ref.\,\cite{kaushal2015spin}. Consequently, Eqs.\,\eqref{eq:weak ion} and \eqref{eq: saddle time} indicate that the time measured by the attoclock can be expressed as the weak value of the temporal delay:
\begin{align}\label{eq:attoweak}
    \tau_{\rm A} = \mathfrak{R}\braket{t}_{\rm w}^{\rm ion}\,.
\end{align}
This formulation generalizes the attoclock concept, making it applicable to other experimental setups. Additionally, it enables the analysis of the attoclock's behavior beyond the saddle-point approximation, providing insights into how tunneling time may be lost. Note that in the case of very strong fields, which lead to very thin barriers, the tunneling process can be modified due to the emergence of an additional pathway. In this pathway, the electron is first reflected at the exit point of the barrier, then reflected a second time from the core, and finally tunnels again \cite{klaiber2022}. We do not consider this channel in our work, as we focus on the regime of relatively thick tunneling barriers.

\section{Comparing the Larmor Clock and the Attoclock for a 1D model}
In this section, we compare the Larmor clock and attoclock within the context of a static one-dimensional ionization model. The Hamiltonian for this model is given by:
\begin{align}\label{eq:toy model}
    \hat{H} = -\frac{\partial^2}{\partial x^2} - \Tilde{\kappa} \delta(x) - F x\,,
\end{align}
where \(\Tilde{\kappa} = \sqrt{2I_{\rm p}}\). 
The initial value problem, starting from the bound state of the delta potential, is solved using the strong-field approximation (SFA) (see Appendix \ref{sec:methods_initialvalue}). 
Figure \ref{fig:Husimi} illustrates the Husimi distribution of the SFA solution. Notably, the classical trajectory of an electron in the electric field \(-F\), originating at the tunnel exit with zero velocity, aligns with the Husimi distribution for \(x \gg 1\) \cite{ivanov2005anatomy}.

\begin{figure}[ht]
    \centering
    \includegraphics[width = \linewidth]{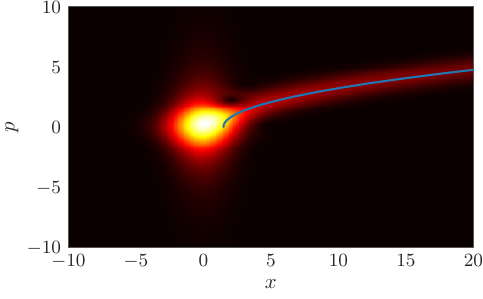}
    \caption{The figure shows the Husimi distribution $|H(x,p)|$ of the SFA solution descibed in \ref{sec:methods_initialvalue} for the ionization potential of Helium and $\kappa = 2$. The blue line shows the classical trajectory of an electron in the electric field, starting with zero velocity at the tunnel exit $x_0$. Far away from the tunnel exit, the classical trajectory coincides with the maximum of the Husimi distribution.}
    \label{fig:Husimi}
\end{figure}

\subsection{Larmor Clock}\label{sec:results larmor}

We now calculate the Larmor time. For the initial state \(\psi_\text{i}\), we use the approximate solution (see \ref{sec:methods_initialvalue},  Eq.\,\eqref{eq:phasetimefunc2} \cite{ivanov2005anatomy}), omitting its global phase factor \(e^{iI_\text{p}t}\). The Fourier transform \(\psi_\text{i}(\xi)\) is computed numerically, where the coordinate \(\xi = x/x_0\) represents the position normalized by the point of exit to the tunnel.

The complex conjugate of the final state corresponds to a particle incident from the right \cite{steinberg1995much}, scattering off the laser potential in the absence of the binding delta potential. Since the potential becomes infinitely large as \(\xi \to -\infty\), the incoming wave will be fully reflected. Consequently, the wave function must vanish as \(\xi\) approaches negative infinity. This implies that the wavefunction can only be represented by the Airy function of the first kind:
\begin{align}
    \psi_{\rm f}(x) = \text{Ai}\left(\kappa^{2/3}(1-\xi)\right)\,.
\end{align}

We calculate the Larmor time using a scaling factor derived in Appendix \ref{sec:scaling_factor}:
\begin{align}
    \tau_\text{L}(x) = \sigma \int_0^{x} dx \, \theta_{\rm B}(x) \psi_\text{f}^*(x)\psi_\text{i}(x)\,.
\end{align}

\begin{figure}[ht]
     \centering
     \begin{subfigure}[b]{1\columnwidth}
         \centering
         \includegraphics[width=\textwidth]{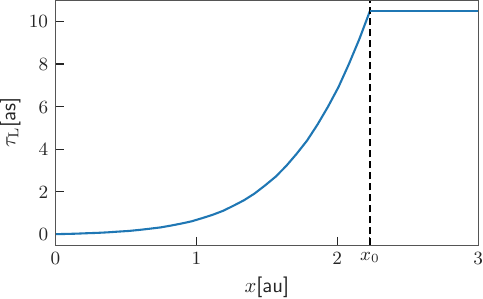}
         \caption{The figure shows the Larmor Time as a function of position calculated using Steinberg's method for the ionization potential of Helium and $\kappa = 3$. The dashed line shows the tunnel exit $x_0$ after which no tunneling time is accumulated by the Larmor Clock.}
         \label{fig:tau_of_x}
     \end{subfigure}
     \hfill
     \begin{subfigure}[b]{1\columnwidth}
         \centering
         \includegraphics[width=\textwidth]{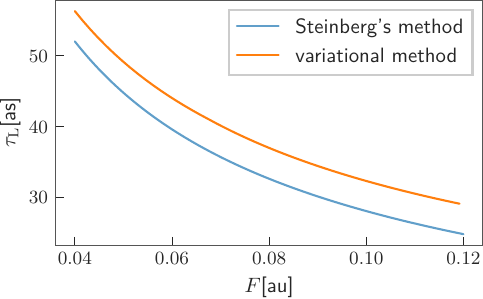}
         \caption{Comparison between Steinberg's method and the variational method of calculating the Larmor time as a function of field strength F. The varitional method is detailed in \ref{sec: variational}.}
         \label{fig:tau_of_F}
     \end{subfigure}
\end{figure}

Figure \ref{fig:tau_of_x} shows the position-resolved Larmor time during tunnel ionization, revealing a non-zero tunneling time. This result is consistent with the findings of Steinberg \cite{steinberg1995much, ramos2020measurement} and Landsman et al. \cite{landsman2014ultrafast, zimmermann2016tunneling}. The accumulated tunneling time increases as the electron moves further from the core. Upon reaching the tunnel exit, the Larmor time stabilizes and remains constant.

Additionally, Figure \ref{fig:tau_of_F} shows the Larmor time as a function of the electric field strength, providing further insight into the tunneling dynamics. The figure compares two methods for calculating the Larmor time. The first method, described in the current paper, follows Steinberg’s approach, explicitly using the initial and final states. The second, variational method, employed by Landsman, is detailed in the appendix \ref{sec: variational}. The two methods exhibit similar behavior, though the variational method consistently shows a slight upward shift. This discrepancy arises because our implementation of Steinberg's method uses  the approximate solution of the time-dependent problem (see \ref{sec:methods_initialvalue}), whereas the variational method is obtained by finding complex energy solutions of the stationary problem (see \ref{sec: variational}).

\subsection{Attoclock}\label{sec:results attoclock}
The ionization amplitude, see Eq.\,\eqref{eq:phasetimefunc} in \ref{sec:methods_initialvalue}, is used to calculate the weak value of temporal delay [see Eq.\,\eqref{eq:attoweak}]:
\begin{align}\label{eq:sok}
    &\tau_{\rm A}= \nonumber\\
    &\text{Re}\left\{\frac{\int\limits_{0}^\infty dt_{\rm D}\, t_{\rm D} e^{-iI_1} e^{-iI_{\rm p}t_{\rm D}} \bra{p - Ft_{\rm D}} \hat{V}_{\rm L} \ket{\phi}}{\int\limits_{0}^\infty dt_{\rm D} e^{-iI_1} e^{-iI_{\rm p}t_{\rm D}} \bra{p - Ft_{\rm D}} \hat{V}_{\rm L} \ket{\phi}} \right\} - \frac{p}{F}\,.
\end{align}
Here, we subtracted an additional term, \(p/F\). The photoelectron amplitude, \(a_{p}\), is defined as the overlap between the solution to the physical problem \(\ket{\psi}\) and the plane waves \(\ket{p}\). Since the electric field is constant in this model, the kinetic momentum, \(p\), is not a "good" quantum number. Consequently, for our static model, Eq.\,\eqref{eq:attoweak} includes an additional contribution due to the constant acceleration in the electric field. This term does not correspond to a tunneling delay and must therefore be subtracted.

By substituting \( u' = (F t_{\rm D} - p) / \sqrt{2 I_{\rm p}} \) into Eq.\,\eqref{eq:sok} and using Eq.\,\eqref{eq:overlap u} for the overlap, we obtain:
\begin{align}\label{eq:attoooo}
    &\tau_{\rm A}(u) = \nonumber\\
    &\tilde{\tau}\text{Re}\left\{\frac{\int\limits_{-u}^{\infty} du' \, u' \exp\left(-i\kappa\left(\frac{u'^3}{3} + u'\right)\right) \frac{u'}{(u'^2 + 1)^2}}{\int\limits_{-u}^{\infty} du' \exp\left(-i\kappa\left(\frac{u'^3}{3} + u'\right)\right) \frac{u'}{(u'^2 + 1)^2}} \right\},
\end{align}

where $\tilde{\tau} = \tilde{\kappa}/F$. As demonstrated in Ref.\,\cite{ivanov2005anatomy}, the classical trajectory of an electron starting at the tunnel exit with zero velocity asymptotically matches the quantum trajectory. This correspondence is evidenced by the Husimi distribution shown in Fig.\,\ref{fig:Husimi}. 

\begin{figure}[ht]
    \centering
    \includegraphics[width = \linewidth]{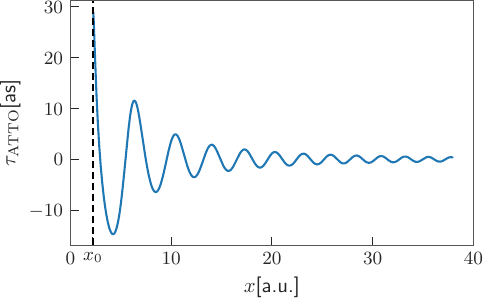}
    \caption{The figure shows the attoclock time as a function of position.}
    \label{fig:attoclocktime}
\end{figure}

Figure \ref{fig:attoclocktime} presents the attoclock weak value, where the classical trajectory of the electron was used to map the momentum into position. The results indicate that the attoclock time is non-zero at the tunnel exit but diminishes as the electron moves away from the atom. This behavior can also be deduced directly from Eq.\,\eqref{eq:attoooo}: as $u\to\infty$, due to parity only the real part in the numerator survives, while only the imaginary part of the denominator remains. Consequently, the real part of the attoclock time vanishes in this limit. In contrast, Fig.\,\ref{fig:tau_of_x} illustrates that the Larmor time does not vanish, proving that the attoclock does not measure the Larmor time.

The Larmor time is a local time \cite{hauge1989tunneling}, meaning that its measurement is spatially confined to the barrier region. In contrast, the attoclock measures an asymptotic time, associated with the electron’s behavior after ionization. The lack of space-resolved measurement erases the information about the Larmor tunneling time from the standard attoclock observables. 

\section{Conclusion}
Previous literature suggested that the attoclock measures the Larmor time \cite{landsman2014ultrafast, zimmermann2016tunneling}. By applying Steinberg’s weak value interpretation of the Larmor clock, we retrieved the position-resolved time density during tunnel ionization, obtaining a non-zero Larmor time consistent with the results of Steinberg \cite{ramos2020measurement, steinberg1995conditional, steinberg1995much} and earlier findings by Landsman et al. \cite{landsman2014ultrafast, zimmermann2016tunneling}. 

We further demonstrated that the attoclock time can be interpreted as the weak value of temporal delay, as introduced by Sokolovski \cite{sokolovski2018no}. While both the Larmor clock and the attoclock can be interpreted through weak values, they correspond to the weak values of two distinct concepts. We show that the weak value of temporal delay, measured by the attoclock, does not coincide with the tunneling time obtained from the Larmor clock. 

Although the attoclock aims to detect the tunneling time at the barrier exit, we have shown that this time vanishes at the momentum-resolved detectors positioned far away from the atom, such as the VMI or COLTRIMS setups used to detect attoclock observables. Formally, this implies that the attoclock does not measure the "local" Larmor time but instead a "non-local" time closely related to the phase time. The absence of a position-resolved measurement, which is crucial for detecting the Larmor time, results in the loss of tunneling time information in the attoclock setup. Consequently, the attoclock fails to resolve the tunneling time at the barrier exit. 

This finding suggests that the tunneling time is not necessarily imprinted onto the electron wavefunction during strong-field processes. Instead, it can only be rigorously measured by detecting the spatial phase of the electron at the barrier exit. Such a measurement could be achieved using a setup similar to the one realized in the BEC tunneling experiment \cite{ramos2020measurement}.

Acknowledgment 
We acknowledge fruitful discussions with P. Stammer.
Funded by the European Union (ERC, ULISSES, 101054696). Views and opinions expressed are however those of the author(s) only and do not necessarily reflect those of the European Union or the European Research Council. Neither the European Union nor the granting authority can be held responsible for them.

\newpage
\bibliographystyle{quantum}
\bibliography{bibliography.bib}

\onecolumn
\appendix

\newpage
\section{Appendix}

\subsection{Attoclock in PPT} \label{sec:spectrum PPT}
In Ref.\,\cite{torlina2015interpreting} numerical analysis showed that the attoclock offset angle vanishes for a short-range binding potential. Here, we aim to show this analytically. For short-range binding potentials, the theory of Perelomov, Popov, and Terent’ev (PPT) \cite{perelomov1966ionization, perelomov1967ionization} provides a suitable framework. According to this theory, the photoelectron spectrum of the attoclock is
\begin{align}
\label{eq:pes}
    |a_{\bm p}(T)|^2 &\propto |\phi_{lm}(\bm{p} + \bm{A}(t_{\rm s}))|^2 e^{2\text{Im}S(T,\bm{p},t_{\rm s})}\,,
    \intertext{with}
    \phi_{lm}(\bm{p}) &= \frac{1}{2}\left[\bm{p}^2 + 2I_{\rm p}\right]\varphi_{lm}(\bm{k})\,,
\end{align}
where $\varphi_{lm}(\bm{k})$ is the wave function of the electron in momentum space.
Let us explicitly calculate the photoelectron spectrum starting from the ground state of a short-range potential
\begin{align}
    \varphi_{00}(\bm{k}) \propto \frac{r_{00}(k)}{k^2 + 2I_{\rm p}}\,,
\end{align}
where the radial function $r_{00}(k)$ has to be calculated numerically. It was shown in Ref.\,\cite{perelomov1966ionization} that 
\begin{align}
    \phi_{00}(\bm{p} + \bm{A}(t_{\rm s})) \propto 1\,,
\end{align}
and thus the photoelectron spectrum is:
\begin{align}\label{eq:spec_simplified}
    |a_{\bm p}(T)|^2 \propto e^{2\text{Im}S(T,\bm{p},t_{\rm s})}\,.
\end{align}
Introducing the vector potential
\begin{align}
    \bm{A} &= -A_0(t) \left[ \cos(\omega t) \bm{e}_x + \sin(\omega t) \bm{e}_y \right]\,
    \intertext{with the envelope}
    A_0(t) &= A_0 \cos^4(\omega t/4)\,,
\end{align}
we can start looking for solutions $t^{\alpha}_{\rm s}$ to the saddle point equation
\begin{align}
    \frac{\partial S(T,\bm{p},t)}{\partial t} \bigg|_{t = t_{\rm s}} = p^2 + A_0^2(t^{\alpha}_{\rm s}) - 2p A_0(t^{\alpha}_{\rm s})\left[\cos\theta \cos\omega t^{\alpha}_{\rm s} +  \sin \theta \sin\omega t^{\alpha}_{\rm s}\right] + 2I_{\rm p} = 0\,, \label{eq: atto_minfunc}
\end{align}
where we have written the momentum $\bm{p}$ in polar coordinates $\{p,\theta\}$. The complex conjugate root theorem states that if $t_{\rm s}^{\alpha}$ is a root of a polynomial, its complex conjugate $(t_{\rm s}^{\alpha})^*$ is also a root. Since every function in Eq.\,\eqref{eq: atto_minfunc} is analytic, the complex root theorem is valid. Furthermore, Eq.\,\eqref{eq: atto_minfunc} remains invariant under the variable transform $(\theta, t_{\rm s}^{\alpha}) \to (-\theta, -t_{\rm s}^{\alpha})$. However, saddle points with negative imaginary components are unphysical, as they produce photoelectron spectra that diverge in the limit $p\to\infty$.

\begin{figure}[!htb]
    \centering
    \includegraphics[width = 0.7\textwidth]{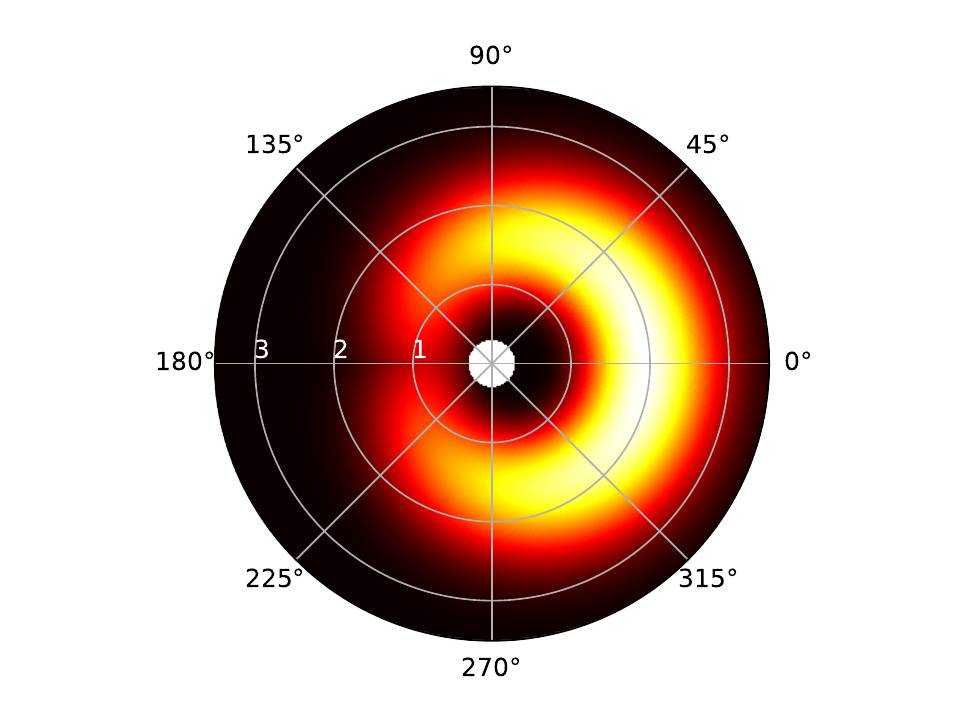}
    \caption{Photoelectron spectrum $|a_{\bm p}|^2$ displayed in polar coordinates $\bm{p} = \{p, \theta\}$, with $p$ in atomic units. The offset angle is zero, indicating a zero attoclock time. The calculation was performed using the ionization potential of Helium and a laser frequency of $0.569$ a.u..}
    \label{fig:spectrum}
\end{figure}

This implies that if $t_{\rm s}^{\alpha} = t_{\rm i}^\alpha + i\tau^\alpha$ is a saddle point for $\theta$, the correct saddle-point for $-\theta$ is $-(t_{\rm s}^{\alpha})^* = -t_{\rm i}^\alpha + i\tau^\alpha$. Let's analyze how the photoelectron spectrum changes when reflecting along the x-axis. We focus on the imaginary part of the action, which is relevant to the photo-electron spectrum in Eq.\,\eqref{eq:pes}:
\begin{align}
    \text{Im}S(T,p, \theta,t_{\rm i} + i\tau) &= \text{Re}\Bigg\{ \frac{1}{2}\int\limits_{\tau}^{0} p^2 + A_0^2(t_{\rm i} + i\tau')\nonumber\\
    &\quad- 2p A_0(t)\left[\cos\theta \cos(\omega(t_{\rm i} + i\tau')) + \sin \theta \sin(\omega(t_{\rm i} + i\tau')) \right] d\tau'\Bigg\} -I_{\rm p}\tau\label{eq:atto_intiterm}\,.
\end{align}
After reflection along the x-axis, the imaginary part of the action becomes:
\begin{align}
    \text{Im}S(T, p, -\theta, -t_{\rm i} + i\tau)  &=\text{Re}\Bigg\{ \frac{1}{2}\int\limits_{\tau}^{0} p^2 + A_0^2(-t_{\rm i} + i\tau') - 2p A_0(-t_{\rm i} + i\tau')\nonumber\\
    &\times\quad\left[\cos\theta \cos(\omega(-t_{\rm i} + i\tau')) - \sin \theta \sin(\omega(-t_{\rm i} + i\tau')) \right] d\tau'\Bigg\} -I_{\rm p}\tau\,.
\end{align}
The first term remains trivially unchanged. The second term is invariant because
\begin{align}
    \text{Re}\cos^4(\omega(-t_{\rm i} + i\tau')) = \text{Re}\cos^4(\omega(t_{\rm i} + i\tau'))\,.
\end{align}
The third term, however, requires a more detailed analysis. Using the identity for real parts of products, $\text{Re}\{ab\} = \text{Re}\{a\}\text{Re}\{b\} - \text{Im}\{a\}\text{Im}\{b\}$, we expand:
\begin{align}
    \text{Re}\left\{A_0(-t_{\rm i} + i\tau')\left[\cos\theta \cos(\omega(-t_{\rm i} + i\tau')) - \sin \theta \sin(\omega(-t_{\rm i} + i\tau')) \right]\right\}\nonumber\\
    = \text{Re}\left\{A_0(-t_{\rm i} + i\tau')\right\} \text{Re}\left\{\left[\cos\theta \cos(\omega(-t_{\rm i} + i\tau')) - \sin \theta \sin(\omega(-t_{\rm i} + i\tau')) \right]\right\}\nonumber\\
    - \text{Im}\left\{A_0(-t_{\rm i} + i\tau')\right\} \text{Im}\left\{\left[\cos\theta \cos(\omega(-t_{\rm i} + i\tau')) - \sin \theta \sin(\omega(-t_{\rm i} + i\tau')) \right]\right\}\label{eq:atto_bloaasd}
\end{align}
Using the following symmetry properties of $\cos$ and $\sin$ under transformations of $t_{\rm i}$:
\begin{align}
    \text{Re}\cos(t_{\rm i} + i\tau') &= \cos(t_{\rm i}) \cosh(\tau') \rightarrow \text{even wrt. }t_{\rm i}\,,\\
    \text{Re}\sin(t_{\rm i} + i\tau') &= \sin(t_{\rm i}) \sinh(\tau') \rightarrow \text{odd wrt. }t_{\rm i}\,,\\
    \text{Im}\cos(t_{\rm i} + i\tau') &= -\sin(t_{\rm i}) \sinh(\tau') \rightarrow \text{odd wrt. }t_{\rm i}\,,\\
    \text{Re}\sin(t_{\rm i} + i\tau') &= \cos(t_{\rm i}) \sinh(\tau') \rightarrow \text{even wrt. }t_{\rm i}\,,
\end{align}
we deduce:
\begin{align}
    \text{Re}\left\{A_0(-t_{\rm i} + i\tau')\left[\cos\theta \cos(\omega(-t_{\rm i} + i\tau')) - \sin \theta \sin(\omega(-t_{\rm i} + i\tau')) \right]\right\}\nonumber\\
    = \text{Re}\left\{A_0(t_{\rm i} + i\tau')\left[\cos\theta \cos(\omega(t_{\rm i} + i\tau')) + \sin \theta \sin(\omega(t_{\rm i} + i\tau')) \right]\right\}\,.
\end{align}
From this, it follows that:
\begin{align}
    S(T, p, \theta, t_{\rm i} + i\tau) = S(T, p, -\theta, -t_{\rm i} + i\tau)\,.
\end{align}
This demonstrates that the photo-electron spectrum is symmetric about $\theta = 0$, resulting in no offset angle. Consequently, the attoclock does not detect a tunneling time for short-range potentials. Fig.\,\ref{fig:spectrum} shows the photo-electron spectrum calculated using Eq. \eqref{eq:spec_simplified} with a single saddle-point for every value of $p$ and $\theta$. We use the saddle point with the smallest positive imaginary $\tau$ value and smallest absolute real value $|t_{\rm i}|$. The figure confirms that the deflection angle is zero, consistent with the analytical findings.

\newpage
\subsection{Weak Value of Temporal Delay in Ionization} \label{sec:attoweak}
Using analytic R-matrix theory (see Ref.\,\cite{torlina2012time}), the ionization amplitude is
\begin{align}
    a_{\bm{p}}(t_{\rm f}) &= \frac{i\tilde{\kappa}}{(2\pi)^{3/2}} \int_{t_0}^{t_{\rm f}}dt\,a_g(t) \int d\bm{r}' e^{-i S(t_{\rm f},t,\bm{r}',\bm{p})} \delta(r'-a) \phi_{\tilde{\kappa},l}(r')Y_{lm}(\theta,\phi)\,,\label{eq:ap arm}
    \intertext{with}
    S(t_{\rm f},t,\bm{r}',\bm{p}) &= \frac{1}{2}\int_{t}^{t_{\rm f}} d\tau \left[\bm{p} + \bm{A}(\tau)\right]^2 +\int_t^{t_{\rm f}}d\tau U[\bm{r}_L(\tau)] + \left[\bm{p} + \bm{A}(t)\right]\cdot \bm{r}' - I_{\rm p} (t-t_0)\,.
\end{align}
Taking the derivative with respect to the ionization potential, while holding $\tilde{\kappa}$ constant, gives
\begin{align}
    \left(\frac{\partial a_{\bm p}}{\partial I_{\rm p}}\right)_{\tilde{\kappa} = \text{const.}} = i (t-t_0) a_{\bm p}\,,
\end{align}
which proves Eq.\,\eqref{eq:weak ion}. Using the saddle-point approximation for the temporal and spacial integrals (see Refs. \cite{kaushal2013nonadiabatic, torlina2013time}) gives
\begin{align}
    a_{\bm p} &= R_{\tilde{\kappa} l m}e^{-iW_{\rm C}(t_{\rm f},\bm{p})}e^{-iS_{\rm SFA}(t_{\rm f},\bm{p})}\,,
    \intertext{where}
    S_{\rm SFA}(t_{\rm f},\bm{p}) &= \frac{1}{2}\int_{t_{\rm s}}^{t_{\rm f}}d\tau \left[\bm{p}+\bm{A}(\tau)\right]^2 - I_{\rm p} t_{\rm s}\,,\\
    W_{\rm C}(t_{\rm f},\bm{p}) &= \int_{t_{\rm s}-\frac{i}{\tilde{\kappa}^2}}^{t_{\rm f}} d\tau U[\bm{r}_{\rm s}(\tau)]\,,\\
    \bm{r}_{\rm s}(\tau) &= \int^{\tau}_{t_{\rm s}} d\tau' \left[\bm{p}+\bm{A}(\tau')\right]\,.
\end{align}
Now, the derivative with respect to the ionization potential, while holding $\tilde{\kappa}$ constant, is
\begin{align}
    \left(\frac{\partial a_{\bm p}}{\partial I_{\rm p}}\right)_{\tilde{\kappa} = \text{const.}} = i (t_{\rm s}-t_0) a_{\bm p}\,,
\end{align}
which proves Eq.\,\eqref{eq: saddle time}.

\subsection{Initial Value Problem and SFA}\label{sec:methods_initialvalue}
We aim to solve the initial value problem starting from the ground state of the delta potential $\phi(x) = e^{-\Tilde{\kappa}|x|}$, with $\Tilde{\kappa} = \sqrt{2I_{\rm p}}$. The time evolution is given by the Hamiltonian
\begin{align}
    \hat{H} &= \hat{V}_\text{L} + \hat{H}_0\,,
    \intertext{with}
    \hat{H}_0 &= -\frac{1}{2} \frac{\partial^2}{\partial x^2} - \Tilde{\kappa} \delta(x), \quad V_\text{L} = -F x\,.
\end{align}
Applying the strong-field approximation yields
\begin{align}
     \psi(v,t) &= -i \int\limits_0^t dt' e^{-i\int\limits_{t'}^t E(t'') dt''}\bra{v(t')} \hat{V}_\text{L} e^{iI_\text{p}t'} \ket{\phi}\,.
\end{align}
Using the relations:
\begin{align}
    E(t'') &= \frac{1}{2}[v(t'')]^2\,,
    \intertext{and}
    v(t'') &= v - A(t) + A(t'')\,,\quad \text{with} \quad A(t) = Ft\,,
\end{align}
we can write:
\begin{align}
    \psi(v,t) &= -i \int\limits_0^t dt' e^{-i\int\limits_{t'}^t \frac{1}{2}[v - F(t-t'')]^2 dt''}\bra{v - F(t-t')} \hat{V}_\text{L} e^{iI_\text{p}t'} \ket{\phi}\,.
\end{align}
Taking a closer look at the integral in the exponent and substituting $\tilde{t} = t - t''$, we have:
\begin{align}
    I_1 &= \int\limits_{t'}^t \frac{1}{2}[v - F(t-t'')]^2 dt''\,,\\
    &=  -\int\limits_{t_\text{D}}^0 \frac{1}{2}[v-F\tilde{t}]^2 d\tilde{t}\,,
\end{align}
where $t_\text{D} = t-t'$ is defined as the delay time. Solving the integral yields
\begin{align}
    I_1 &=  \frac{t_\text{D}}{6} ( 3v^2-3Fvt_\text{D}+t_\text{D}^2F^2)\,.
\end{align}
Substituting $t_\text{D} = t-t'$, the wave function now is
\begin{align}
    \psi(v,t) &= -i e^{iI_\text{p}t} \int\limits_{t}^0 dt_\text{D} e^{-iI_1} e^{-iI_\text{p}t_\text{D}} \bra{v - Ft_{\rm D}} \hat{V}_\text{L}  \ket{\phi}\label{eq:phasetimefunc}
\end{align}
Next, substituting $u' = (F t_\text{D} - v) / \tilde{\kappa}$:
\begin{align}\label{eq:sys bubbel}
    \psi(u,t) &= -i \tilde{\tau} e^{iI_\text{p}t} e^{-i\kappa\left(\frac{u^3}{3}+u\right)} \int\limits_{t/\tilde{\tau}-u}^{-u} du' e^{-i\kappa\left(\frac{u'^3}{3}+u'\right)} \braket{-\tilde{\kappa}u' | \hat{V}_\text{L} | \phi}\,.
\end{align}
We have defined $\tilde{\tau} = \Tilde{\kappa}/F$, $u = v/\Tilde{\kappa}$, and $\kappa = I_{\rm p}\Tilde{\kappa} / F$. 
We can now consider the matrix element in the integrand of Eq.\,\eqref{eq:sys bubbel}:
\begin{align}
    \braket{-\tilde{\kappa}u' | \hat{V}_\text{L} | \phi} = -F \int\limits_{-\infty}^\infty dx\,e^{ix\tilde{\kappa}u'} x e^{-\tilde{\kappa}|x|} = F \frac{x_0^2}{i\kappa^2} \frac{4u'}{(u'^2+1)^2}\,. \label{eq:overlap u}
\end{align}
The classical turning point is $x_0 = I_{\rm p}/F$. Finally, the whole function in the infinite time limit yields a stationary solution
\begin{align}\label{eq:phasetimefunc2}
    \psi(u,t\to\infty)  &= \frac{4x_0}{\kappa} e^{iI_\text{p}t} e^{-i\kappa\left(\frac{u^3}{3}+u\right)} \int\limits_{-u}^{\infty} du' e^{-i\kappa\left(\frac{u'^3}{3}+u'\right)} \frac{u'}{(u'^2+1)^2}\,.
\end{align}

\subsubsection{Full Saddle Point Approximation}
To derive an analytical solution for the wave function in Eq.,\eqref{eq:phasetimefunc2}, we apply the saddle-point method. The saddle points are determined by solving the saddle-point equation:
\begin{align}
    \frac{\partial}{\partial u'} i\kappa\left(\frac{u'^3}{3}+u'\right) &= 0\\
    \Leftrightarrow u'^2 + 1 &= 0 \\
    \Rightarrow u'_{1/2} &= \pm i\,.
\end{align}
We will only consider the even part of the integrand, since for the odd part the integration limits can always be switched, leaving the saddle point outside the integration region:
\begin{align}
    \psi(u) = -i \frac{4 x_0}{\kappa} e^{-i\kappa\left(\frac{u^3}{3} + u\right)} \underbrace{\int\limits_{-u}^\infty du'\, \frac{u'}{(u'^2+1)^2} \sin\left(\kappa\left(\frac{u'^3}{3} + u'\right)\right)}_{A=}\,.
\end{align}
The variable transformation $y = u' + u$ and yields
\begin{align}
    A = \int\limits_{0}^\infty \frac{(y-u)}{\left(\left(y-u\right)^2+1\right)^2}\sin\left(\kappa\left(\frac{(y-u)^3}{3}+y-u\right)\right) dy\,.
\end{align}
To use the saddle point approximation, we can rewrite $A$ as
\begin{align}
    A = \text{Im}\left\{ \int\limits_{0}^\infty \overbrace{\frac{(y-u)}{\left(\left(y-u\right)^2+1\right)^2}}^{f(y)}\exp\left(i\overbrace{\kappa\left(\frac{(y-u)^3}{3}+y-u\right)}^{S(y)}\right) dy \right\}\,.
\end{align}
The slowly varying function $f(y)$ has a pole at the saddle points $y_{1/2} = u\pm i$, where $S'(y_{1/2}) = 0$. Nonetheless, we can use the asymptotic solution provided in \cite{long2010keldysh}. We rewrite $A$ as
\begin{align}
    A = \int\limits_{0}^\infty \underbrace{\frac{(y-u)}{\left(y-y_2\right)^2}}_{g(y)} \frac{1}{\left(y-y_1\right)^2} \exp\left(\kappa \underbrace{i \left(\frac{(y-u)^3}{3}+y-u\right)}_{w(y)}\right)\,,
\end{align}
and use the approximation
\begin{align}
    {A} &\approx- \pi i g(y_1) e^{\kappa w(y_1)} \left(\frac{w''(y_1)}{2}\right)^\frac{1}{2} \frac{1}{\Gamma\left(\frac{3}{2}\right)} \sqrt{\kappa}\,.
\end{align}
With 
\begin{align}
    g(y_1) = -\frac{i}{4}\,, \quad w(y_1) = -\frac{2}{3}\,, \quad w''(y_1) = -2\,,
\end{align}
we get 
\begin{align}
    A \approx-  \text{Im}\left\{\pi i \frac{-i}{4} e^{-\frac{2}{3}\kappa} \sqrt{-1} \sqrt{\kappa} \frac{2}{\sqrt{\pi}}  \right\} = -\text{Im}\left\{i \frac{\sqrt{\kappa\pi}}{2} e^{-\frac{2}{3}\kappa} \right\} = -\frac{\sqrt{\kappa\pi}}{2} e^{-\frac{2}{3}\kappa}\,.
\end{align}
This expression holds for $u>0$, because only then the saddle point is within the integration interval. This means that one could argue $A\approx 0 + \mathcal{O}(1/\sqrt{\kappa})$ for $u<0$, allowing us to write
\begin{align}
    A = -\left( \theta(u) \frac{\sqrt{\kappa\pi}}{2} e^{-\frac{2}{3}\kappa} + \mathcal{O}\left(\frac{1}{\sqrt{\kappa}}\right) \right)\,.
\end{align}
The wave function then is
\begin{align}\label{eq: psi theta}
    \psi(u) \approx 2 i x_0 \sqrt{\frac{\pi}{\kappa}} e^{-\frac{2}{3}\kappa} e^{-i\kappa\left(\frac{u^3}{3} + u\right)} \theta(u)\,.
\end{align}
Its Fourier transform is
\begin{align}
    \psi_\text{S}(\xi) &\approx i\frac{2}{\sqrt{2\pi}} x_0 \sqrt{\frac{\pi}{\kappa}} e^{-\frac{2}{3}\kappa} \int\limits_{0}^\infty du\, e^{-i\kappa\left(\frac{u^3}{3}+(1-\xi)u\right)}\nonumber\\
    &= i \pi\sqrt{2} x_0 \kappa^{-5/6} e^{-\frac{2}{3}\kappa} \left[\text{Ai}(\kappa^{2/3}(1-\xi)) - i \text{Gi}(\kappa^{2/3}(1-\xi))\right]\,.\label{eq:LTIon_FTapprox}
\end{align}
For $\xi\to\infty$, where $\text{Bi}(\kappa^{2/3}(1-\xi)) \simeq \text{Gi}(\kappa^{2/3}(1-\xi))$ (see \cite{abramowitz1988handbook}), the wave function asymptotically becomes
\begin{align}\label{eq:fullsaddleS}
    \psi_\text{S}(\xi) &\simeq \sqrt{2} i \pi x_0 \kappa^{-5/6} e^{-\frac{2}{3}\kappa} \left[\text{Ai}(\kappa^{2/3}(1-\xi)) - i \text{Bi}(\kappa^{2/3}(1-\xi))\right]\,.
\end{align}

\subsection{Scaling Factor} \label{sec:scaling_factor}
In scattering, the weak value of the projector $\hat{\theta}_{\rm B}$ is divided by the incident flux, ensuring that the Larmor time has units of time. However, in the case of strong-field ionization, it is not possible to define an incident flux. To get units of time in strong-field ionization, we introduce a scaling factor
\begin{align}\label{eq:scaling_const}
    \sigma = \lim_{x\to\infty} \frac{1}{v(x) \Psi_\text{f}^*(x) \Psi_\text{i}(x)}\,,
\end{align}
where $v(x)$ is the classical velocity of the particle. With this scaling factor, the Larmor time becomes
\begin{align}
    \tau_\text{L} = \sigma \int\limits_{a}^{b}dx\,\psi_\text{f}^*(x)\psi_\text{i}(x)\,.\label{eq:lc_lt_scaling}
\end{align}
This ensures that the asymptotic Larmor velocity coincides with the classical velocity. To compute $\sigma$, we need the asymptotic expressions for $\xi \to\infty$ of the initial and final states as $\xi \to \infty$. Using the asymptotic solution given by Eq.\,\eqref{eq:fullsaddleS} and the classical velocity in a constant fieldm $v(\xi) = \sqrt{2I_\text{p}\xi}$, the scaling factor is:
\begin{align}
    \sigma^{-1} &= \lim_{\xi\to\infty} \sqrt{2I_\text{p}\xi}\, \psi_\text{f}^*(\xi) \psi_\text{i}(\xi)\,.
\end{align}
Assuming $\psi_{\rm i}(\xi) \simeq \psi_{\rm S}(\xi)$ as $\xi\to\infty$, we have
\begin{align}
    \sigma^{-1} &= \lim_{\xi\to\infty} \sqrt{2I_\text{p}\xi}\, A \left[A_\text{R}^* \text{Ai}(\chi)+ B_\text{R}^* \text{Bi}(\chi)  \right] \left[\text{Ai}(\chi) - i \text{Bi}(\chi)\right]\,,
\end{align}
where $\chi \coloneqq \kappa^{2/3}(1-\xi)$. Using the asymptotic forms for $\text{Ai}(\chi)$ and $\text{Bi}(\chi)$ from \cite{abramowitz1988handbook}, we rewrite the scaling factor as:
\begin{align}
    \sigma^{-1} &= \lim_{\xi\to\infty} \sqrt{2I_\text{p}\xi}\frac{A}{2\pi\sqrt{-\chi}} \left[ A_\text{R}^* \left(1-e^{2i\left(\frac{2}{3}(-\chi)^{3/2}+\frac{\pi}{4}\right)}\right) - iB_\text{R}^* \left(1+e^{2i\left(\frac{2}{3}(-\chi)^{3/2}+\frac{\pi}{4}\right)}\right) \right]
\end{align}
By neglecting oscillatory terms and using $\sqrt{-\chi} \sim \sqrt{\kappa^{2/3} \xi}$, we find
\begin{align}
     \sigma &= \frac{2^{2/3}\pi}{F^{1/3} A \left(A_\text{R}^* - iB_\text{R}^*\right)}\,. \label{eq:scalingF}
\end{align}

\subsection{Variational Method for Computing Larmor Time in 1D Tunneling} \label{sec: variational}

Here, we outline the method used to compute the Larmor time for a one-dimensional model of tunnel ionization with a time-independent electric field. This method was used by Zimmermann et al. in Ref. \cite{zimmermann2016tunneling}.

The Larmor time is computed on the basis of the phase evolution of the wave function's transmission coefficient as a function of a small perturbation of the potential. The Larmor time is then given by:
\begin{align}
    \tau = -\frac{\partial \theta}{\partial V}\bigg|_{E},
\end{align}
where $\theta$ is the phase of the transmission coefficient, and $V$ is the perturbation of the potential and the energy $E$ is kept constant.

The stationary Schrödinger Equation is
\begin{align}
    \left( -\frac{1}{2} \frac{\partial^2}{\partial x^2} - \tilde{\kappa} \delta(x) - F x + dV\theta_{\rm B}(x) - E \right) \psi = 0\,
\end{align}
The wave function is divided into three regions:
\begin{align}
    \psi(x) = 
    \begin{cases} 
        \psi_\text{L}(x) = A_\text{L} \text{Ai}\left(s + \frac{x}{l}\right) & \text{for } x < 0, \\ 
        \psi_0(x) = A_0 \text{Ai}\left(s_0 + \frac{x}{l}\right) + B_0 \text{Bi}\left(s_0 + \frac{x}{l}\right) & \text{for } 0 \leq x < x_0, \\ 
        \psi_\text{R}(x) = A_\text{R} \left(\text{Ai}\left(s + \frac{x}{l}\right) - i \text{Bi}\left(s + \frac{x}{l}\right)\right) & \text{for } x \geq x_0,
    \end{cases}
\end{align}

with the parameters defined as:
\[
s = -\frac{E}{\beta}, \quad s_0 = -\frac{E - dV}{\beta}, \quad l = -\frac{\beta}{F}, \quad \beta = \left(\frac{F^2}{2}\right)^{1/3}.
\]

Up to normalization, the transmission coefficient is
\begin{align}
    T = \psi(x_0) = \psi_\text{R}(x_0)
\end{align}

Setting $A_\text{L} = 1$ and enforcing continuity of the wave function and its derivative gives the following system of linear equations

\begin{align}
    \begin{pmatrix}
        \text{Ai}(s_0) & \text{Bi}(s_0) & 0 \\
        \text{Ai}'(s_0) & \text{Bi}'(s_0) & 0 \\
        \text{Ai}\left(s_0 + \frac{x_0}{l}\right) & \text{Bi}\left(s_0 + \frac{x_0}{l}\right) & \text{Ai}\left(s + \frac{x_0}{l}\right) - i\text{Bi}\left(s + \frac{x_0}{l}\right) \\
        \text{Ai}'\left(s_0 + \frac{x_0}{l}\right) & \text{Bi}'\left(s_0 + \frac{x_0}{l}\right) & \text{Ai}'\left(s + \frac{x_0}{l}\right) - i\text{Bi}'\left(s + \frac{x_0}{l}\right)
    \end{pmatrix}
    \begin{pmatrix}
        A_0\\
        B_0\\
        A_\text{R}
    \end{pmatrix}
    =
    \begin{pmatrix}
        \text{Ai}(s) \\
        \text{Ai}'(s) - 2 \tilde{\kappa} l \, \text{Ai}(s) \\
        0 \\
        0
    \end{pmatrix}.
\end{align}
To determine the complex energy eigenvalues, $dV$ is set to zero and an iterative optimization process is employed. The energy is treated as a complex variable, and its value is refined by minimizing a residual function that quantifies the mismatch in the boundary conditions. The eigenfunction $\psi_{dV = 0}(E_0,x)$ with the longest lifetime is chosen. The derivative w.r.t. $V$ while holding $E_0$ constant is taken numerically by minimizing the residual function for $dV$ being small but non-zero, yielding $\psi_{dV = 0}(E_0,x)$. The Larmor time then is
\begin{align}
    \tau = - \frac{\arg\left(\psi_{dV}(E_0,x)\right) - \arg\left(\psi_{dV=0}(E_0,x)\right)}{dV}
\end{align}

\subsection{Connection between the variational method and Steinberg's method}\label{sec: connection}
Let $\psi$ be a solution
\begin{align}
    i \frac{\partial}{\partial t} \ket{\psi} = H \ket{\psi}\,.
\end{align}
And $\psi_{dV}$ be a solution
\begin{align}
    i \frac{\partial}{\partial t} \ket{\psi_{dV}} = \left( H +dV \hat{\theta}_{\rm B} \right) \ket{\psi_{dV}}\,,
\end{align}
where
\begin{align}
    H = \frac{\hat{p}^2}{2} + V(x)\,.
\end{align}
We can write the varied wave funciton as
\begin{align}
    \ket{\psi_{dV}(t)} = - i \int_{t_0}^t e^{- i \int_{t'}^{t} H d\tau} dV \hat{\theta}_{\rm B} \ket{\psi_{dV}(t')} dt' + e^{- i \int_{t_0}^{t} H d\tau} \ket{\psi_{dV}(t_0)}\,.
\end{align}
Let $\ket{\psi(t_0)} = \ket{\psi_{dV}(t_0)}$ then we can rewrite as
\begin{align}
        \ket{\psi_{dV}(t)} = - i \int_{t_0}^t e^{- i \int_{t'}^{t} H d\tau} dV \hat{\theta}_{\rm B} \ket{\psi_{dV}(t')} dt' + \ket{\psi(t)}\,.
\end{align}
The transmission amplitude is obtained by projecting onto the the final state
\begin{align}
    T = \braket{f|\psi_{dV}}\,.
\end{align}
The Larmor time obtained by the variational method is
\begin{align}
    \tau_{\rm L} = i \frac{\partial}{\partial \theta} \ln T = \frac{T'}{T} i\,.
\end{align}
Using
\begin{align}
    i T' = \frac{\braket{f|\psi_{dV}}-\braket{f|\psi}}{dV} = \int_{t_0}^t \bra{f}e^{- i \int_{t'}^{t} H d\tau} \hat{\theta}_{\rm B} \ket{\psi_{dV}(t')} dt'
\end{align}
and
\begin{align}
    \psi_{dV}(t') \approx \psi(t')
\end{align}
we get
\begin{align}
    \tau_{\rm L} = \frac{\int_{t_0}^t \braket{f(t') | \hat{\theta}_{\rm B} | \psi(t')} dt'}{\braket{f(t) |\psi(t)}}\,.
\end{align}
If initial and final states are not time-dependent, we retrieve
\begin{align}
    \tau_{\rm L} = \frac{\braket{f | \hat{\theta}_{\rm B} | \psi} }{\braket{f |\psi}}\,.
\end{align}

\subsubsection{Example in Scattering}
To illustrate the equivalence of Steinberg's method and the variational method, we consider the scattering of a potential barrier with classical turning points at $-x_0$ and $x_0$.
The solution for scattering of a barrier for a particle incident from the left is
\begin{align}\label{eq:initial}
    \psi_{i}(x<-x_0) = e^{ikx} + R e^{-ikx}\,, \quad \psi_{i}(x>x_0) = T e^{ikx}\,,
\end{align}
representing the initial state in Steinberg's formulation \cite{steinberg1995much}.
Following Ref. \cite{steinberg1995much}, the final state representing transmission is the complex conjugate of the solution of a particle incident from the right:
\begin{align}
    \psi_{t}^*(x<-x_0) = T_{t} e^{-ikx}\,, \quad \psi_{t}^*(x>x_0) = R_{t} e^{ikx} + e^{-ikx}\,.
\end{align}
The coefficients of the initial and transmitted states are connected by the Wronskian relations \cite{sokolovski1987traversal}:
\begin{align}
    T = T_{t} \,,\quad R T^* +R_{t}^* T = 0\,.
\end{align}
Using these, the transmitted state can be reformulated as:
\begin{align}\label{eq:transmitted}
    \psi_{t}^*(x<x_0) = T e^{-ikx}\,, \quad \psi_{t}^*(x>x_0) = -\frac{R^*T}{T^*} e^{ikx} + e^{-ikx}\,.
\end{align}
The final state representing reflection is simply the complex conjugate of the initial state \cite{steinberg1995much}:
\begin{align}
    \psi_{r}(x) = \psi_{i}^*(x)\,.
\end{align}
The initial state can be expanded in terms of reflected and transmitted parts:
\begin{align}\label{eq:split}
    \ket{i} = \braket{r|i}\ket{r} + \braket{t|i}\ket{t} = R \ket{r} + T \ket{t}\,.
\end{align}
Introducing a small perturbation $\epsilon \hat{\theta}_{\rm B}$, where $\epsilon = \tau \tilde{\epsilon} \ll 1$, to the initial state under the barrier gives:
\begin{align}
    \ket{i_\epsilon} &= e^{-i\epsilon\hat{\theta}_{\rm B}} \ket{i}.
\end{align}
Expanding this:
\begin{align}
    \ket{i_\epsilon} &\approx R \ket{r} \left(1- i\epsilon \frac{\braket{r|\hat{\theta}_{\rm B}|\psi}}{\braket{r|\psi}}\right) + T \ket{t} \left(1- i\epsilon \frac{\braket{t|\hat{\theta}_{\rm B}|\psi}}{\braket{t|\psi}}\right)\,\label{eq:blaal2}
\end{align}
Next, we calculate the Larmor time using the variational method. The derivative of the perturbed initial state w.r.t. $\epsilon$ is
\begin{align}
    i \psi_i'(x>x_0) &= R \psi_r(x>x_0) \frac{\braket{r|\hat{\theta}_{\rm B}|\psi}}{\braket{r|\psi}} + T \psi_t(x>x_0) \frac{\braket{t|\hat{\theta}_{\rm B}|\psi}}{\braket{t|\psi}}\\
    &= \frac{\braket{t|\hat{\theta}_{\rm B}|\psi}}{\braket{t|\psi}} Te^{ikx} + \left(\frac{\braket{r|\hat{\theta}_{\rm B}|\psi}}{\braket{r|\psi}} - \frac{\braket{t|\hat{\theta}_{\rm B}|\psi}}{\braket{t|\psi}}  \right) RT^* e^{-ikx} \label{eq:blaaal}
\end{align}
The second term represents a left-going solution and vanishes for a symmetric barrier. Furthermore, solving the varied Schrödinger equation for a particle incident from the left implicitly projects out the second term. Thus the derivative reduces to:
\begin{align}
    i \psi_i'(x>x_0) &= \frac{\braket{t|\hat{\theta}_{\rm B}|i}}{\braket{t|i}} Te^{ikx}
\end{align}
From this, the Larmor time is:
\begin{align}
    \tau_{\rm L} = -\frac{\partial\theta}{\partial V} = i \frac{\psi_i'(x>x_0)}{\psi_i(x>x_0)} = \frac{\braket{t|\hat{\theta}_{\rm B}|i}}{\braket{t|i}}\,.
\end{align}
This result demonstrates the equivalence between Steinberg's method and the variational method when the variation is performed under the correct boundary conditions.

\end{document}